\documentclass[12pt]{article}
\def\vep{\varepsilon}
\def\al{\alpha}

\def\beq{\begin{equation}}
\def\eeq{\end{equation}}
\def\pd{\partial}
\def\bea{\begin{eqnarray}}
\def\eea{\end{eqnarray}}
\begin{document}
\begin{center}
{\Large{\bf Integration measure and 
 extended BRST  covariant  quantization }}\\
\vspace{0.5cm}
{\large Bodo Geyer}$^{a}$\footnote{e-mail: geyer@itp.uni-leipzig.de},
{\large Petr Lavrov}$^{a,b}$\footnote{e-mail: lavrov@tspu.edu.ru}and
{\large Armen Nersessian}$^{a,c,d}$\footnote{e-mail: nerses@thsun1.jinr.ru}
\end{center}
\vspace{0.5cm}
{\it $^{a}$ Center of Theoretical Studies, Leipzig University,
Augustusplatz 10/11, D-04109 Leipzig, Germany\\
 $^{b}$ Tomsk State Pedagogical University,
634041 Tomsk, Russia\\
$^{c}$ JINR, Laboratory of Theoretical Physics,  Dubna,  141980 Russia  \\
$^{d}$ Yerevan State University, A. Manoogian St., 3,
Yerevan, 375025, Armenia} 
\begin{abstract}
We propose an extended BRST invariant Lagrangian quantization scheme   
of general gauge theories  based on explicit realization 
of the "modified triplectic algebra" that was announced in our previous 
investigation~(hep-th/0104189). The algebra includes, besides 
the specific odd operators $V^a$ appearing 
in the triplectic formalism, also the odd operators $U^a$ introduced
within modified triplectic quantization and the second-order odd
operators $\Delta^a$. While the operators $V^a$    
can be viewed as anti-Hamiltonian vector fields generated by
a second-rank irreducible $Sp(2)$ tensor,
the operators $U^a$  are the anti-Hamiltonian 
vector fields generated by a $Sp(2)$ scalar.
We show that some even supersymplectic structure defined on 
the space of fields and antifields, provides the 
extended BRST path integral with a well-defined integration measure.
All the known Lagrangian  quantization schemes 
based on the extended BRST symmetry are obtained by specifying the
(free) parameters of that method.   
\end{abstract}
\renewcommand{\thefootnote}{\arabic{footnote}}
\setcounter{footnote}0
\thispagestyle{empty}
\setcounter{equation}0
\section{Introduction}

The Batalin-Vilkovisky (BV) formalism \cite{BV} of Lagrangian 
quantization of general gauge theories, since its introduction, 
attracts permanent interest due to its covariance, universality 
and mathematical elegance.  The BV formalism 
is outstanding also from a pure mathematical point of view
because it is formulated in terms of seemingly exotic objects: the
antibracket (odd Poisson bracket) and the (related) second-order operator 
$\Delta$. The study of the formal geometrical structure of the BV formalism,
performed during the last ten years, allowed to introduce its 
interpretation in terms of more traditional mathematical objects 
 as well as to find the unusual behaviour of the 
antibracket with respect to integration theory \cite{bvgeom}.

On the other hand, there exists an extended,
$Sp(2)$ symmetric (BRST/antiBRST invariant) version of the
BV formalism \cite{BLT}, and its geometrized version  
known as ``triplectic formalism'' \cite{BM,BMS} (see also \cite{BM1,ND,GGL}). 

The main ingredients of triplectic quantization are a pair of
antibrackets $(\cdot,\cdot)^a$, a pair of operators $\Delta^a$, and
a pair of odd vector fields $V^a$, with $a=1,2$ being related to
BRST and antiBRST symmetry. However,
the degeneracy of the antibrackets appearing in the $Sp(2)$ symmetric
quantization schemes, and the structure of $V^a$ fields, lead to
difficulties in establishing the links between the various
quantization schemes as well as in the geometrical analysis of their 
structure \cite{ND}.
For example, within the ``canonical triplectic quantization'' 
\cite{BM,BMS} the initial version of the $Sp(2)$ symmetric Lagrangian
quantization \cite{BLT} is not contained. However, within the ``modified 
triplectic quantization'' \cite{GGL} which is specified by the presence 
of an additional pair of odd operators $U^a$ the initial $Sp(2)$ symmetric 
formalism \cite{BLT} is contained as a special limit. 
Thereby, the separation of appropriate variables [in our context to be
identified with $x=\{\phi, \bar\phi\}$]
can be considered as essential means for a consistent 
quantization procedure allowing for the formulation of a suitable boundary 
condition for quantum action and for the definition of an appropriate 
Lagrangian manifold on which a natural volume form is defined ensuring the
integral of the vacuum functional well-defined. 
The pair of operators $U^a$ appearing in the modified triplectic 
quantization \cite{GGL} may serve for these purposes. 

An interesting  observation on the  structure of $Sp(2)$ invariant formalism
was made in Ref.~\cite{BMS}: there, it was found that the 
``triplectic algebra'' generates a Poisson bracket on an appropriate 
subspace of the whole space of fields and antifields.
However,  this Poisson bracket did not appear in  the triplectic formalism.
It seemed, that it plays an auxiliary role, giving the means for the
 geometrical formulation  of the gauge-fixing conditions, and the 
compatibility conditions for the antibrackets and $V^a$ fields as well 
\cite{ND}.

In our recent  paper \cite{abp}, starting from this observation, 
we proposed a covariant realization of the triplectic algebra 
 by making use of this Poisson bracket and of a flat connection
which respects it.
We have shown, that the operators $V^a$    
can be viewed as the anti-Hamiltonian vector fields generated by
a second-rank irreducible $Sp(2)$ tensor, $S_{ab}$,
while the operators $U^a$  are the anti-Hamiltonian 
vector fields generated by some  $Sp(2)$ scalar, $S_0$. 
 There,
it was also found that the whole space of the triplectic formalism can
be equipped with an even symplectic structure.
Note, that for the first time a flat symplectic connection was 
used by I.~A.~Batalin and I.~V.~Tyutin for the covariant formulation 
 of deformation quantization in Ref.~\cite{BT}.

Now,  in this paper we define the extended BRST invariant path integral
whose integration measure corresponds to that symplectic structure, and 
we formulate the master equations for the quantum action $W$ and 
gauge-fixing
functional $X$. Furthermore, we show that the 
various known
Lagrangian quantization schemes based on the extended BRST symmetry 
\cite{BLT,BM,GGL} are included
in the former one as special cases.
Hence,  the above-mentioned Poisson bracket 
 plays the  basic role in the  $Sp(2)$ symmetric Lagrangian quantization 
schemes:  It provides them with a well-defined integration measure and
allows for a reparametrization-invariant realization of 
triplectic algebra.

The paper is arranged as follows. 

In Section 2 we give a brief description of the general statements and 
 explicit realizations of the triplectic quantization scheme. 

 In Section 3,  for the reason of completeness,
 the basic ingredients of quantization scheme including 
 the $\Delta^a-$operators, the antibrackets $(\,\cdot\,,\,\cdot\,)^a$,
 the $V^a$- and $U^a$- fields   are presented. These
 objects were proposed in \cite{abp} and given in reparametrized invariant
 form with the help of a flat symmetric
 connection which respects the Poisson bracket on the space of fields. 
 
In Section 4 we complete these structures by
the integration measure  defined by the use of an even symplectic 
structure which is constructed on 
the whole space of fields and antifields.   
 Then we construct a quantization scheme based on the 
 modified triplectic algebra and we show that all the known approaches 
 to triplectic quantizations can be considered as specific ones within 
 the proposed method. 

Finally in the Appendix we collect various relations among these objects 
which we need for the construction of the triplectic algebra.

\setcounter{equation}{0}
\section{Extended BRST symmetric formalisms}

Let us present the basic ingredients of extended BRST invariant Lagrangian
quantization schemes, thereby, following the original papers \cite{BLT,BM,BMS}.

In these schemes one requires the existence of a pair of odd (differential) 
operators $\Delta^a$ and of odd vector fields $V^a$ ($a = 1, 2$) 
satisfying the following consistency conditions:
\begin{eqnarray}
&\Delta^{\{a}\Delta^{b\}} =0,& \quad \epsilon(\Delta^a)=1, \label{dd} \\
&V^{\{a}V^{b\}} =0,&\quad \epsilon(V^a)=1, \label{vv}\\
&\Delta^{\{a} V^{b\}} + V^{\{a}\Delta^{b\}} = 0. &
\label{vdw}
\end{eqnarray}
Here, and in the following, the curly bracket denotes symmetrization with
respect to the enclosed indices $a$ and $b$.
The operators $\Delta^a$, due to their obstruction of the Leibniz rule,
generate a pair of antibrackets,
\beq
(-1)^{\epsilon(f)}(f,g)^a = \Delta^a(fg) - (\Delta^af) g
 -(-1)^{\epsilon(f)}f(\Delta^a g) ,
\label{liebdelta} 
\eeq
which obey the properties of graded antisymmetry,
\beq
\label{antisym}
(f,g)^a=-(-1)^{(\epsilon(f)+1)(\epsilon(g)+1)}
(g,f)^a,
\eeq 
of the Leibniz rule,
\beq
\label{Lr}
(f,gh)^a=(f,g)^ah+(-1)^{\epsilon(h)\epsilon(g)}(f,h)^ag,
\eeq
and of the consistency conditions:
\begin{eqnarray}
& &(-1)^{(\epsilon(f)+1)(\epsilon(h)+1)}(f,(g,h)^{\{a})^{b\}} + ~
{\mbox{\rm cycl. perm.}}~ = 0, \quad \label{bj}\\
& &\Delta^{\{a}(f,g)^{b\}} = (\Delta^{\{a} f, g)^{b\}}
+(-1)^{\epsilon(f)+1}(f,\Delta^{\{a} g)^{b\}} ~,    \label{ad}\\
& &V^{\{a}(f,g)^{b\}} = (V^{\{a}f,g)^{b\}}  
+(-1)^{\epsilon (f)+1} (f, V^{\{a} g)^{b\}}. \label{avw}
\end{eqnarray}

The partition function in the triplectic formalism
is defined by the expression
\begin{equation}
Z=\int dz \; d\lambda \;{\rm exp}\big\{(i/\hbar)\left[W + X\right]\big\}~,
\label{z}
\end{equation}
where $z$ denotes the whole set of fields and antifields,
$W=W(z)$ is viewed as the quantum action of the theory, 
and $X=X(z,\lambda)$ is considered as the gauge fixing term. Obviously,
this division of the gauge fixed action into two pieces is to a large 
extent arbitrary, but we shall follow the conventions of Refs.~\cite{BM,BMS}.
The gauge-fixing function $X$ restricts the partition function to the 
``space
 of effective fields'' being necessary to describe the quantum dynamics, and
$\lambda$ are a some additional parametric field variables which
simply become 
Lagrangian multipliers of gauge constraints when $X$ depends on them only
linearly.

The partition function (\ref{z}) is gauge independent if the following
quantum master equations hold:
\begin{equation}
(\Delta^a + \hbox{\Large$\frac{i}{\hbar}$}V^a)
\exp\big\{\hbox{\Large$\frac{i}{\hbar}$}W\big\} = 0~,
\qquad
(\Delta^a-\hbox{\Large$\frac{i}{\hbar}$}V^a + \ldots)
\exp\big\{\hbox{\Large$\frac{i}{\hbar}$}X\big\} = 0~.
\label{2m}
\end{equation}
In the last equation the dots indicate those extra terms which are required
due to the variation of the  field $\lambda$. The explicit form of the 
second master equation will be specified in the following.  
The extended BRST transformations are generated by the operators $\delta^a$,
\beq
\label{brstanti}
 \delta^a=(X -W,\;\cdot\;)^a+2V^a +i\hbar\Delta^a~.
\eeq

For convenience let us now give the explicit realizations of the 
triplectic algebra and the corresponding quantization procedure in the 
various versions of the extended BRST invariant Lagrangian formalism
being under consideration in the literature. 
%
%
\subsection{$Sp(2)-$symmetric version \cite{BLT}}

In the initial version of the $Sp(2)$ invariant formalism the triplectic 
algebra is realised on the ``physical fields'' $\phi^A$ (including the 
original fields, ghosts, antighosts and Lagrangian multipliers), 
the (supplementary) fields $\bar\phi_A$ of the same grading 
$\epsilon (\bar\phi_A)=\epsilon(\phi^A)\equiv\epsilon_A$ and the pair of
antifields $\phi^{*}_{Aa}$ of the opposite grading
$\epsilon(\phi^{*}_{Aa})=\epsilon_A +1.$

On these fields the triplectic algebra is realized as follows
($\phi^{*a}_A\equiv\vep^{ab}\phi^*_{Ab},\; \vep^{ab}=-\vep^{ba}$):
\beq
\label{DaVa}
 \Delta^a  =  \Delta_0^a 
 \equiv 
 (-1)^{\epsilon_A}\frac{\partial_l}{\pd\phi^A}
 \frac{\partial}{\partial\phi^{*}_{Aa}},
 \qquad 
 V^a  =  {\hat V}^a_0
 \equiv  
 \vep^{ab}\phi^{*}_{Ab}\frac{\partial}{\partial\bar\phi_A},
\eeq
where, according to Eq.~(\ref{liebdelta}), 
the corresponding antibrackets are defined by the relations
\beq
 (\phi^A,\phi^{*}_{Bb})^a=\delta^a_b\delta^A_B.
\eeq

In order to be able to formulate the gauge fixing conditions 
in Ref.~\cite{BLT}, via introduction of additional fields $\pi^{Aa}$, 
$\epsilon(\pi^{Aa})=\epsilon_A+1$, and $\lambda^A$, 
$\epsilon(\lambda^A)=\epsilon_A$, 
a further extension of the space of fields has been performed. 
In these terms the partition function can be presented in the form
\beq
\label{Z}
 Z=\int dz\; d\lambda \; {\rm exp}\left \{\hbox{\Large$\frac{i}{\hbar}$}(W+
 X +S_0)\right \}
\eeq
 where $z^{\mu}=(\phi^A,\phi^*_{Aa};{\bar\phi}_A,\pi^{Aa})$,
 $W=W(\phi,\phi^*,\bar\phi)$ is the quantum action, and
 the functionals $X=X(\phi,\pi,\bar\phi;\lambda)$
and $S_0=S_0(\phi^*,\pi)$ are introduced as follows:
\beq
X=\left(\bar\phi_A-
\frac{\partial G}{\pd\phi^A}\right)\lambda^A
-\hbox{\Large$\frac{1}{2}$}\vep_{ab}\pi^{Aa}\frac{\pd^2G}{\pd\phi^A\pd\phi^B}
\pi^{Bb},
\eeq
\beq
\label{S0}
S_0=\phi^*_{Aa}\pi^{Aa};
\eeq
the functional $G=G(\phi)$ defines a specific choice of the gauge.

In this version the quantum action $W$ is supposed to obey
the master equation (\ref{2m}) together with the boundary condition
to coincide with the initial classical action when 
$\hbar = \phi^* = \bar\phi = 0$; no condition is required for $X$.

\subsection{Triplectic version \cite{BM}}

 Another version of the  $Sp(2)$ formalism, suggested by Batalin and 
 Marnelius~\cite{BM}, considers the auxiliary fields $\pi^{aA}$ as 
 canonically conjugate with respect to the fields $\bar\Phi_A$.
 The triplectic algebra is realized in the following way:
\bea
\label{trV}
 \Delta^a=\Delta_0^a +(-1)^{\epsilon^A+1}\vep^{ab}
 \frac{\partial_l}{\partial\pi^{Ab}}\frac{\pd}{\partial\bar\phi_A},
 \quad 
V^a=\hbox{\Large$\frac{1}{2}$}\Big({\hat V}^a_0+
(-1)^{\epsilon_A+1}\pi^{Aa}\frac{\partial_l}{\pd\phi^A}\Big),
\eea
with $\Delta_0^a$ and ${\hat V}^a_0$ being defined through 
Eqs.~(\ref{DaVa}); then, the corresponding antibrackets are defined 
by the relations
\beq
 (\phi^A,\phi^*_{Bb})^a=\delta^a_b\delta^A_B,
\qquad
 (\bar\phi_A,\pi^B_b)^a=\delta^a_b\delta^B_A,
\eeq
thus making obvious the triplet structure of the formulation.
For later convenience let us write the $V^a-$operators as
\beq
\label{Ua}
V^a=\hbox{\Large$\frac{1}{2}$}({\hat V}^a_0+{\hat U}_0^a)
\qquad
{\rm with}
\qquad
{\hat U}_0^a\equiv
(-1)^{\epsilon_A+1}\pi^{Aa}\frac{\partial_l}{\pd\phi^A}.
\eeq
 
While the quantum action $W=W(z)$ is required to satisfy (\ref{2m}),
the gauge fixing functional 
$X=X(z,\lambda)$, 
is required to satisfy the following master equation:
\beq
\label{X}
\left(\Delta^a - 
\hbox{\Large$\frac{i}{\hbar}$}V^a\right){\rm exp}\{(i/\hbar)X \}=0.
\eeq
Notice that by the explicit realization of the anti-Hamiltonian
vector fields $V^a$ in the form (\ref{trV}), cf., also Eq.~(\ref{Vec1}), 
and of the above introduced antibrackets the stronger relations
\beq
\label{VD}
\Delta^a V^b+ V^a\Delta^b=0,
\eeq
\beq
\label{Vanti}
V^a(f,g)^b=(V^a f,g)^b-(-1)^{\epsilon(f)}(f,V^ag)^b
\eeq
hold, i.e., without symmetrization of the indices $a$ and $b$. 

\subsection{Modified triplectic version \cite{GGL}}

 The essential point of the original triplectic quantization \cite{BM}
consists in
 dividing the task of constructing the partition function $Z$ into two parts:
 first, in the construction of the quantum action $W$, and second, in the
 construction of a suitable gauge fixing functional $X$. Either problem is
 solved by means of the appropriate master equations (\ref{2m}), (\ref{X}).
 Unfortunately, the explicit realization of this attractive idea within 
 original version meets the problem of the boundary condition for $W$. Due to 
the special structure of the operators $V^a$, Eq.~(\ref{trV}), 
 it is not possible, in contrast to all
 previously known schemes of Lagrangian quantization, to consider the
 initial classical action as a natural boundary condition to the solution
 of the quantum master equations. This was the main reason 
 in Ref.~\cite{GGL} to reformulate the original triplectic scheme by a 
 modified one.  

 Remaining in the same configuration space
 of fields and antifields $z=(\phi,\phi^*;{\bar\phi},\pi)$, and 
 accepting the idea of a separate treatment
 of the two above mentioned actions $W$ and $X$, it was 
 proposed to change from the beginning the triplectic algebra 
 and the generating master equations by introducing an additional set of 
 $Sp(2)$ doublets of operators $U^a$, $\epsilon(U^a)=1$, by requiring
\begin{eqnarray}
\Delta^{\{a}\Delta^{b\}} =0, \quad V^{\{a}V^{b\}} =0, 
\quad\Delta^{\{a} V^{b\}} + V^{\{a}\Delta^{b\}} = 0,
\end{eqnarray}
\begin{eqnarray}
\label{mal}
U^{\{a}U^{b\}}=0,\quad \Delta^{\{a} U^{b\}} + U^{\{a}\Delta^{b\}} = 0,
\quad U^{\{a}V^{b\}} +V^{\{b}U^{a\}}=0.
\end{eqnarray}
This algebra can be considered as an extension of triplectic algebra
(\ref{dd}), (\ref{vv}) and (\ref{vdw}) and is refered to as the "modified 
triplectic algebra". 
An explicit realization of the operators may be given by
$\Delta^a$ as introduced in Eqs.~(\ref{trV}), 
$V^a \equiv {\hat V}^a_0$ and 
$U^a \equiv {\hat U}^a_0$, as introduced in Eqs.~(\ref{DaVa}) 
and (\ref{Ua}), respectively. 

Again, the partition function is given by the functional integral (\ref{Z}),
\beq
\label{Zm}
 Z=\int dz\;d\lambda \; {\rm exp}\left\{(i/\hbar)(W + X +S_0)\right \},
\eeq
but now $W=W(z)$ and $X=X(z,\lambda)$ are required to satisfy the 
following quantum master equations:
\beq
 \left(\Delta^a + 
 \hbox{\Large$\frac{i}{\hbar}$} V^a\right){\rm exp}\{(i/\hbar) W\} =0,
\qquad
 \left(\Delta^a - 
 \hbox{\Large$\frac{i}{\hbar}$} U^a\right){\rm exp}\{(i/\hbar) X \}=0,
\eeq
where the functional $S_0$ was defined in (\ref{S0}).
 Now one can use the standard boundary condition for $W$ 
 in the form of the initial classical action
 when $\hbar = \phi^* = \bar\phi = 0$. 
Note that the difference of the operators
$V^a-U^a\equiv{\hat V}^a_0-{\hat U}^a_0$ can be represented as  
anti-Hamiltonian vector fields: 
\beq
V^a-U^a=(S_0,\;\cdot\;)^a .
\eeq

\setcounter{equation}0
\section{Triplectic algebra, modified triplectic algebra and Poisson brackets}

The basic ingredients  of the triplectic formalism and the quantization 
in arbitrary coordinates were formally constructed  and analyzed 
in terms of an ``anti-triplectic non-degenerate metric'' \cite{BMS}. 

An important feature of triplectic algebra which was observed in
 Ref.~\cite{BMS} is the existence of some Lagrangian subspace
${\cal M}_0$ 
on which a $Sp(2)$ invariant even Poisson bracket
can be defined, viz.,
\beq
 \{u,w\}_0 \equiv \epsilon_{ab}(u, V^aw)^b ~.
\label{even}\eeq
This subspace ${\cal M}_0$ is specified by requiring
for any functions $u$, $v$, $w$ on ${\cal M}_0$ to hold:
$$(u,v)^a = 0,\qquad  (u,V^{\{a} v)^{b\}},w)^c=0.$$

In the canonical and primary triplectic versions \cite{BLT,BM} of the
$Sp(2)$ covariant Lagrangian quantization schemes
this construction yields a canonical Poisson bracket
(in Darboux coordinates $x=\{\phi,\bar\phi\}$):
\beq
\{u(\phi,\bar\phi),v(\phi,\bar\phi)\}_0=
\frac{\partial_r u}{\partial \phi^A}
\frac{\partial_l v}{\partial \bar\phi_A}
-
\frac{\partial_r u}{\partial \bar\phi_A}
\frac{\partial_l v}{\partial\phi^A}.
\eeq

Now, we try to find an explicit realization of the triplectic
algebra on a Poisson (super)space
where the Poisson bracket $\{\;\cdot\;,\;\cdot\;\}_0$ 
in local coordinates $x^i$ (``fields'') is given by the
expression
\beq
\{u(x), v(x)\}_0=\frac{\partial_r u(x)}{\partial x^i}
\omega^{ij}(x)\frac{\partial_l v(x)}{\partial x^j},
\label{pbx}
\eeq
where
\beq
\omega^{ij}=-(-1)^{\epsilon_i\epsilon_j}\omega^{ji},
\qquad 
(-1)^{\epsilon_i\epsilon_k}
\omega^{kl}\pd_l\omega^{ij}\;+
\; {\rm cycl.~perm}\;(ijk)=0. \label{pbe}
\eeq

Then we consider the superspace ${\cal M}$ parametrized by the
coordinates  $z^{\mu} =(x^i,\;\theta_{ia})$, 
($\epsilon(\theta_{ia})=\epsilon_i+1$), 
where $\theta_{ia}$ (''antifields'') are 
transformed as $\pd_i=\partial/\partial x^i$ 
under reparametrizations of  ${\cal M}_0$.~\footnote{
For simplicity, in the following 
we restrict ourself to the case when all coordinates $x^i$ are even,
$\epsilon_i=0$, although in the context of quantum field theory among 
$x^i$ odd variables (ghost fields) always are present. The transition to 
the general case can easily be performed.
}  
This superspace can be equipped by a pair of ``canonical''
antibrackets $(\;\cdot\;,\;\cdot\;)^a_{\rm can}$,
\beq
\label{antib1}
 (f,g)^a_{\rm can}=\frac{\pd f}{\pd x^i}
 \frac{\partial_l g}{\partial \theta_{ia}}
 +(-1)^{\epsilon(f)}\frac{\partial_l f}{\partial \theta_{ia}}
 \frac{\pd g}{\pd x^i}~.
\eeq  
On that superspace ${\cal M}$ there exists a naturally
defined object being a $Sp(2)$ irreducible second rank tensor, namely
\beq
\label{Sab}
 S_{ab}=\hbox{\Large$\frac{1}{6}$}
\theta_{ia}\omega^{ij}(x)\theta_{jb}~, 
\qquad 
 S_{ab}=S_{ba}~.
\eeq
This tensor obeys the condition
\beq
\label{ji}
(S_{a\{b},S_{c\}d})^d_{\rm can}=0,
\eeq
which is  nothing but the Jacobi identity for the Poisson 
bracket (\ref{pbe}).
With the help of this tensor we introduce the vector
fields ${V}^a$ in an explicitly  $Sp(2)$ symmetric way,
\beq
\label{Vec1}
 V_a=(S_{ab},\;\cdot\;)^b_{\rm can}, 
\eeq
with the following results:
\beq
\label{Vec2}
V^a=\hbox{\Large$\frac{1}{2}$}
\omega^{ij}\theta_j^a \pd_i+
 \hbox{\Large$\frac{1}{6}$}
\pd_i\omega^{jk}\,\theta^a_j\theta_{kb}
 \frac{\pd}{\pd\theta_{ib}},
\qquad \theta^{ia}=\vep^{ab}\theta^i_b.
\eeq
In accordance with (\ref{even}) these vector fields $V^a$ generate 
on ${\cal M}_0$ the even Poisson bracket.

 It should be noticed that it is not necessary to use the algebraic 
properties of $V^a$ which are dictated by the triplectic algebra in 
order to find the relation between antibrackets and Poisson bracket.  
Namely, from the definition (\ref{Vec2}) we obtain the following result
\begin{eqnarray}
\nonumber
V_{\{a}V_{b\}}
&=&(S_{c\{a},(S_{b\}d},\;\cdot\;)^d_{\rm can})^c_{\rm can}=
\\
&&=\hbox{\Large$\frac{1}{12}$}
\omega^{ij}\,\pd_i\omega^{mn}\,\theta_{ma}\theta_{nb}\pd_j+
\hbox{\Large$\frac{1}{6}$}
\left(\hbox{\Large$\frac{1}{6}$}
\pd_i\omega^{mk\,}\pd_j\omega^{in}+
\hbox{\Large$\frac{1}{6}$}
\omega^{km}\pd_i\pd_j\omega^{in}+\right .\\
\nonumber 
&&\left .+\hbox{\Large$\frac{1}{3}$}
(\pd_i\omega^{mn}\,\pd_j\omega^{ik}+
\pd_i\omega^{nk}\,\pd_j\omega^{im}
+\pd_i\omega^{km}\,\pd_j\omega^{in})\right)
\theta_{mb}\theta_{ka}\theta_{nc}\frac{\pd}{\pd\theta_{jc}}.
\label{Veca}
\end{eqnarray}
Obviously, the conditions (\ref{vv}) are failed in general. It can be easily
checked that the conditions (\ref{avw}) are failed as well. 
{\it Hence,  the existence of a $Sp(2)$ invariant Poisson
 bracket defined by the expression} (\ref{even})   {\it does not require 
the   conditions} (\ref{vv}), (\ref{avw}).

To obtain the explicit realization  of the triplectic algebra
generating the Poisson structure (\ref{pbx}), let us 
equip  ${\cal M}_0$ by a symmetric  connection which
respects the Poisson bracket
\beq
\label{sc}
 \Gamma^k_{ij}=\Gamma^k_{ji},
\qquad
\pd_i\omega^{kj}+\omega^{kl}\Gamma^j_{il}+\omega^{lj}\Gamma^k_{il}=0.
\eeq
Notice, that from the symmetry of the Christoffel symbols 
the Jacobi identity for the Poisson bracket follows.
 When the Poisson bracket is
 non-degenerate the symmetric Poisson connection coincides with the 
symmetric symplectic connection which  is known as
 ``Fedosov connection''~\cite{fm} because of Fedosov's work 
 on globally defined deformation quantization \cite{F}. It was  recently 
found to 
be the natural object  in the Hamiltonian BRST quantization 
as well \cite{BGL}. In what follows we consider the case when the Poisson
bracket is non-degenerate. 

Using this   connection, 
we can define  on the superspace ${\cal M}$
the differential operators  
corresponding to the covariant derivative on ${\cal M}_0$:
\beq
\nabla_i=\pd_i +\Gamma^k_{ij}\theta_{ka}\frac{\pd}{\pd\theta_{ja}},\qquad
[\nabla_i,\nabla_j]=R^k_{\;mij}\theta_{ka}\frac{\pd}{\pd\theta_{ma}},
\label{nabla}
\eeq
where $R^k_{\;mij}$ are the components of the curvature 
tensor of the Fedosov connection,
\beq
\label{R}
 R^l_{\;ijk}=\pd_j\Gamma^l_{ki}- \pd_k\Gamma^l_{ij} + 
 \Gamma^m_{ki}\Gamma^l_{jm}-\Gamma^m_{ij}\Gamma^l_{km},
\qquad
 R^l_{\;ijk}= -R^l_{\;ikj}.
\eeq 

To get the realization of the triplectic algebra in general coordinates, one
can consider a minimal generalization by replacing  
within the antibrackets (\ref{antib1}) the  usual derivative, $\pd_i$,
by the covariant one, $\nabla_i$, namely 
\beq
\label{antib}
 (f,g)^a=(\nabla_if)\frac{\pd g}{\pd\theta_{ia}}+(-1)^{(\epsilon(f)}
 \frac{\pd f}{\pd\theta_{ia}}(\nabla_ig)
\eeq
and in the definition of the operators $\Delta^a$, 
\beq
\label{Da1}
\Delta^a=\nabla_i\frac{\pd}{\pd\theta_{ia}} 
+\hbox{\Large$\frac{1}{2}$}(\rho(x),\;\cdot\;)^a~
\eeq
(where $\rho$ is some functional on ${\cal M}_0$ which will be
restricted even more in the Appendix).
These operators obviously generate through (\ref{liebdelta}) a pair of
bilinear operations (\ref{antib}) acting on ${\cal M}$ and obeying
evidently the properties (\ref{antisym}) and (\ref{Lr}).

Taking into account that the following  equation holds:
\beq
\nabla_iS_{ab}=0,
\eeq
we obtain that the vector 
fields $V^a$, defined  through (\ref{antib}), are of the form
\beq
\label{Vec}
V_a=(S_{ab},\;\cdot\;)^b =\hbox{\Large$\frac{1}{2}$}
\theta_{ia}\omega^{ij}\nabla_j.
\eeq
 For these objects to be antibrackets,
these operations (\ref{antib}) must satisfy the Jacobi
identities (\ref{bj}).
On the other hand, the Jacobi identities hold,
if the commutators of the $\Delta^a-$operators are  first order 
differential operators. However,
the explicit calculation yields the result (see Appendix)
\beq
 \Delta^{\{a}\Delta^{b\}}=R^k_{\;lij}\theta_{kc}
 \frac{\pd}{\pd\theta_{lc}}\frac{\pd}{\pd\theta_{ia}}
 \frac{\pd}{\pd\theta_{jb}}.
\eeq
Hence, we should require  $\Gamma^k_{ij}$ to be a flat connection:
  \beq
R^k_{\;ijm}=0.
\eeq 
 The existence of such
flat connections directly follows from the Darboux theorem,
since in Darboux coordinates one can choose the trivial connection which is
obviously flat. Note that every non-linear  canonical transformation
transfers the trivial connection into a non-zero one. Such a flat connection,
respecting the Poisson bracket, was used in Ref.~\cite{BT} for the 
formulation of a coordinate-free scheme of deformation quantization. 

Straightforward calculations
(see  Appendix) of the  algebra of operators 
$\Delta^a$ (\ref{Da1}), vector fields
$V^a$ (\ref{Vec}) and operations $(\;\cdot\;,\;\cdot\;)^a$ yield
the conclusion that the remaining relations of the triplectic algebra
(\ref{dd}), (\ref{vv}), (\ref{liebdelta}), (\ref{bj}), (\ref{ad}),
(\ref{Vanti}) and (\ref{VD}) in its strong version are fulfilled
when  $R^k_{\;ijm}=0$.
Therefore {\it we arrive at an explicit realization of the triplectic
algebra on ${\cal M}$ with an arbitrary flat Poisson space
${\cal M}_0$}.

Now we extend the triplectic algebra to the modified one by the
introduction of additional vector fields $U^a$ as follows: 
Let us equip  ${\cal M}_0$ by a symmetric tensor $g_{ij}(x)=g_{ji}(x)$. 
Then we are able to construct a $Sp(2)$ scalar function $S_0$ being 
defined on the superspace $\cal M$, 
\beq
 S_0=\hbox{\Large$\frac{1}{2}$}\theta_{ia}g^{ij}\theta_{jb}\vep^{ab},
 \qquad \epsilon(S_0)=0, \qquad g^{ij}=\omega^{im}g_{mn}\omega^{nj}.
\eeq
It  generates the anti-Hamiltonian vector fields $U^a$
\beq
\label{Ua1}
U^a=(S_0,\;)^a =\hbox{\Large$\frac{1}{2}$} g^{jl}_{;i}\theta_{jc}\theta^c_l
\frac{\pd}{\pd\theta_{ia}} + g^{ij}\theta^a_j\nabla_i,
\eeq 
where 
\[
g^{jl}_{;i}=\pd_ig^{jl}+\Gamma^j_{in}g^{nl}+ \Gamma^l_{in}g^{jn} 
\]
denotes the covariant derivative of tensor $g^{jl}$. 
The requirement of the conditions (\ref{mal}) yields the following equations
for $S_0$:
\beq
(S_0,S_0)^a=0, \qquad V^aS_0=0,\qquad \Delta^a S_0=0.
\label{S1}
\eeq

Let us define the
traceless matrix $I^i_k$ by the relations 
$I^i_k=\omega^{ij}g_{jk}$.
In terms of the operator $\hat I$ the relations (\ref{S1}) read
\beq
I^k_{j;i}-I^k_{i;j}=0,\qquad
N^k_{ij}\equiv I^l_iI^k_{j;l} -I^l_jI^k_{i;l}-
I^k_l(I^l_{j;i}-I^l_{i;j})=0,
\eeq  
where  $N^k_{ij}$ is the Nijenhuis tensor. When these equations are
fulfilled the set of operators $\Delta^a$, Eq.~(\ref{Da1}), $V^a$, 
Eq.~(\ref{Vec}), and $U^a$, Eq.~(\ref{Ua1}), form the modified triplectic 
algebra (\ref{mal}).

Notice that in Darboux coordinates the matrix $\omega_{ij}$ is 
in the form
\begin{eqnarray*}
\omega_{ij}=
\left( \begin{array}{c} 0\,\,\,\,\,\,\,\,\,{\rm id}\\-{\rm id}
\,\,\,\,\,\,0\end{array},
\right)
\end{eqnarray*}
where `${\rm id}$' is the unit matrix. Then the vector fields $V^a$ (\ref{Vec})
are exactly reduced to the operators used in the original
triplectic quantization \cite{BM},
\begin{eqnarray}
\label{Vad}
V^a=\hbox{\Large$\frac{1}{2}$}({\hat V}^a_0+{\hat U}^a_0),
\end{eqnarray}
where ${\hat V}^a_0$ and
${\hat U}^a_0$ were introduced in (\ref{DaVa}) and (\ref{Ua})
respectively.

Choosing for the  matrix  $g_{ij}$ the representation 
$$
g_{ij}=
\left( \begin{array}{c} 0\,\,\,\,\,\,\,\,{\rm id}\\{\rm id}\,\,\,\,\,\,0
\end{array}
\right), 
$$
we obtain for the vector fields $U^a$ (\ref{Ua1}) the expressions
\begin{eqnarray}
\label{Uad}
U^a={\hat V}^a_0-{\hat U}^a_0. 
\end{eqnarray}
The vector fields used in formulation of modified
triplectic quantization \cite{GGL} are linear combinations of the vector
fields $V^a$ (\ref{Vad}) and $U^a$ (\ref{Uad}). 

\setcounter{equation}0
\section{Quantization procedure}

The superspace ${\cal M}$
can be equipped with an {\it even symplectic structure}.
Indeed we can construct the even closed two-form, 
which is non-degenerated on antifields (cf.,~Ref.~\cite{KN}): 
\beq
\Omega_2=d(\theta_{ia}\omega^{ij} D\theta^a_j)= 
\hbox{\Large$\frac{1}{2}$} R^{.}_{ijkl}\theta^{ak}\theta^l_a dx^i\wedge dx^j
+\omega^{ij}D\theta_{ia}\wedge D\theta^a_j~,
\eeq
where $D\theta_{ia}=d\theta_{ia}-\Gamma^k_{ij}\theta_{ka}dx^j$, 
and $\Gamma^k_{ij}$  is some Fedosov   connection (not necessarily flat).
Hence, requiring the connection to be flat,
we can equip the superspace ${\cal M}$,
in addition to the triplectic algebra,
with an  even symplectic structure 
\beq
\label{ss}
\Omega=dz^\mu\Omega_{\mu\nu}dz^\nu=\omega_{ij}dx^i\wedge dx^j+
\kappa^{-1}\Omega_2=\omega_{ij}dx^i\wedge dx^j+
\kappa^{-1}\omega^{ij}D\theta_{ia}\wedge D\theta^a_j~,
\eeq
where $\kappa$ is an arbitrary constant.
The   corresponding {\it non-degenerate Poisson bracket} reads
\beq
 \{f(z), g(z)\}=(\nabla_i f)\omega^{ij}( \nabla_j g)+
\kappa\frac{\partial_r f}{\partial \theta^a_{i}}
\omega_{ij}\frac{\partial_l g}{\partial \theta_{ja}}~.
\label{pbn}
\eeq

Using the even symplectic structure we can define on 
${\cal M}$ the  analogue of the Liouville measure \cite{KGS},
\beq
{\cal D}_0={\sqrt{
{\rm Ber}\;\Omega_{\mu\nu}}}=
\kappa\left({{\det\;\omega_{ij}}}\right)^{3/2}~,
\label{d0}
\eeq
which
is invariant under supercanonical transformations 
of the Poisson bracket (\ref{ss}).

Now we
construct the vacuum functional and the generating
master equations for a quantum action $W=W(z)$  and a gauge 
fixing functional 
$X=X(z,\lambda)$ using basic operators $\Delta^a, V^a, U^a$ and the function
 $S_0$  introduced above. Define the vacuum 
functional $Z$ as the following path integral
\begin{eqnarray}
\label{Z6}
Z=\int dz \; d\lambda  \;{\cal D}_0\;{\rm exp}
{\{(i/\hbar)[W+X+\al S_0]\}}
\end{eqnarray}
where ${\cal D}_0$ is the integration measure (\ref{d0}) and $\alpha$ is 
a constant. We suggest the following form
of generating master equations for $W$ and $X$,
\begin{eqnarray}
\label{MEW}
\hbox{\Large$\frac{1}{2}$} (W,W)^a +{\cal V}^a W=i\hbar\Delta^a W,
\end{eqnarray}     
\begin{eqnarray}
\label{MEX}
\hbox{\Large$\frac{1}{2}$} (X,X)^a +{\cal U}^a X=i\hbar\Delta^a X,
\end{eqnarray}
where ${\cal V}^a$ and ${\cal U}^a$ are first order differential operators 
constructed from $V^a$ and $U^a$. 

Let us consider the more general transformations of coordinates $z$ 
generated by antibrackets $(\;\cdot\;,\;\cdot\;)^a$ and operators $V^a, U^a$
\begin{eqnarray}
\label{extBRST}
\delta z=(z,F)^a\mu_a + \beta\mu_a V^az + \gamma\mu_a U^a z
\end{eqnarray}
where $\beta,\gamma$ are some constants, $\mu_a$ is a $Sp(2)$
doublet of anticommuting constant parameters and $F=F(z,\lambda)$ is a
bosonic functional.
The transformations (\ref{extBRST}) lead to the variation
\begin{eqnarray}
\nonumber
\delta(W+X+\al S_0)=\mu_a\left[(W+X,F)^a -\al U^aF
+\beta V^a(W+X)+\gamma U^a(W+X)\right].
\end{eqnarray}   
The corresponding Jacobian $J$ has the form
\begin{eqnarray}
\nonumber
J={\rm exp}{\{2\Delta^aF\mu_a\}}.
\end{eqnarray}
Choosing $F=X-W$ and taking into account the generating master equations
(\ref{MEW}),(\ref{MEX}), the requirement of invariance of the integrand  
in (\ref{Z6}) takes the form
\begin{eqnarray}
\nonumber
[2{\cal V}^aW-2{\cal U}^aX-(\al U^a+\beta V^a+\gamma U^a)W+
(\al U^a -\beta V^a-\gamma U^a)X]\mu_a=0.
\end{eqnarray}
These equations are satisfied if we define the operators 
${\cal V}^a,\;{\cal U}^a$ by the relations
\begin{eqnarray}
\label{VU}
{\cal V}^a=\hbox{\Large$\frac{1}{2}$}(\al U^a+\beta V^a+\gamma U^a),\qquad
{\cal U}^a=\hbox{\Large$\frac{1}{2}$}(\al U^a -\beta V^a-\gamma U^a).
\end{eqnarray}
Evidently for arbitrary constants $\al, \beta, \gamma$ the operators
${\cal V}^a,\;{\cal U}^a$ obey the properties
\begin{eqnarray}
{\cal V}^{\{a}{\cal V}^{b\}}=0,\qquad
{\cal U}^{\{a}{\cal U}^{b\}}=0,\qquad
{\cal V}^{\{a}{\cal U}^{b\}}+{\cal U}^{\{a}{\cal V}^{b\}}=0.
\end{eqnarray}
Therefore, the operators $\Delta^a$, Eq.~(\ref{Da1}), and 
${\cal V}^a,\;{\cal U}^a$, Eqs.~(\ref{VU}) realize the modified 
triplectic algebra.
Obviously, the integrand of vacuum functional (\ref{Z6}) is invariant under 
the BRST and antiBRST transformations of the coordinates $z$ 
defined by the generators
\begin{eqnarray}
\label{BRSTg}
 \delta^a=(X-W,\;\cdot\;)^a +{\cal V}^a-{\cal U}^a.
\end{eqnarray} 
Note that the same result concerning the integration measure
${\cal D}_0$ in the functional integral (\ref{Z6}) can be
obtained by requiring the extended BRST symmetry principle.

Finally, we arrive at the conclusion, that the given construction 
includes all the $Sp(2)-$covariant quantization schemes listed in 
Section 2:
\begin{itemize}
\item  $\al=0$:

The operators ${\cal V}^a, \;{\cal U}^a$ are linearly
dependent ${\cal U}^a =-{\cal V}^a$. In this case there exists only one set 
of the vector fields ${\cal V}^a$ under consideration and the modified
triplectic algebra is reduced to the triplectic one. 

\item $\al=0,\beta=2,\gamma=0$:

In this case one obtains ${\cal V}^a= V^a$ in fact
representation of vector fields used in constructing the triplectic
quantization in general coordinates \cite{BMS}. 
In Darboux coordinates the vector fields ${\cal V}^a= \frac
12({\hat V}^a_0+{\hat U}^a_0)$,
${\cal U}^a =-{\cal V}^a$ coincide with ones used in original version of
 the triplectic quantization \cite{BM}.

\item $\al=1,\beta=2,\gamma=0$:

Using Darboux coordinates in the operators 
(\ref{Vad}), (\ref{Uad}) we reproduce the vector field 
${\cal V}^a= {\hat V}^a_0$, 
${\cal U}^a=-{\hat U}^a_0$ adopted in \cite{BLT} 
as well as in the modified triplectic quantization \cite{GGL}.
 In this case $\rho=const$, 
$S_0=\phi^*_{Aa}\pi^{Aa}$ and the vacuum functional (\ref{Z6}) coincides
with that within the modified triplectic quantization (\ref{Zm}).

\end{itemize}

{\it In conclusion, we developed the extended BRST invariant quantization 
scheme in Lagrangian formalism which includes  
all the ingredients  appeared  in realization of the modified triplectic
algebra. The formulation  exactly uses the integration measure 
extracted from the even 
symplectic structure on the whole space of fields and antifields.
It allows   to  consider all existing covariant quantization schemes
with extended BRST symmetry  as special limits.}

\vspace{0.5cm}

\noindent
{\sc Acknowledgements:}~
 We thank A. Karabegov  and I.V. Tyutin for useful discussions. 
 P.L. and A.N. acknowledge the hospitality of NTZ at the Center of
 Advanced Study of Leipzig University and financial support
 by the Saxonian  Ministry of Fine Art and Sciences.
 The  work of P.L. was also supported under the projects
 RFBR 99-02-16617, INTAS 99-0590, DFG-RFBR 99-02-04022 and the 
 Fundamental Sciences Grant E00-3.3-461
 of the Russian Ministry of Education. 
 The work of A.N. was supported under  the INTAS project 00-0262.  

\vspace{0.8cm}
\noindent
{\bf \Large  Appendix}

\appendix
\section{Algebra of operators $\Delta^a, V^a$ and $(\;,\;)^a$}
\setcounter{equation}{0}

Let us consider algebra of operators $\Delta^a$ (\ref{Da1}), $V^a$
(\ref{Vec}) and the operations $(\;,\;)^a$ (\ref{antib}). 
Straightforward calculations yield the following results
\beq
 \Delta^{\{a}\Delta^{b\}}={\hat R}^{ab},\quad 
 {\hat R}^{ab}=R^k_{\;lij}\theta_{kc}
 \frac{\pd}{\pd\theta_{lc}}\frac{\pd}{\pd\theta_{ia}}
 \frac{\pd}{\pd\theta_{jb}},
\eeq
\beq
 V^{\{a}V^{b\}}=\frac 14R^k_{\;mij}\theta^{ia}\theta^{jb}\theta_{kc}
 \frac{\pd}{\pd\theta_{mc}},
\eeq
\begin{eqnarray}
\label{DaVb}
&& 2(\Delta^aV^b+V^b\Delta^a)=
 \vep^{ab}\omega^{ij}R^k_{\;lji}\theta_{kc}\frac{\pd}{\pd\theta_{lc}} +
 R^k_{\;lij}\theta_{kc}\theta^{ib}\frac{\pd}{\pd\theta_{ja}}
 \frac{\pd}{\pd\theta_{lc}}+\\
\nonumber
&&+\vep^{ab}(\pd_i\omega^{ij} -\hbox{\Large$\frac{1}{2}$}\omega^{ji}
\frac{\pd\rho}{\pd x^i})\nabla_j +
 (\pd_i\Gamma^i_{jl}-\Gamma^i_{jm}\Gamma^m_{il} + 
 \hbox{\Large$\frac{1}{2}$}\frac{\pd\rho}{\pd x^i}\Gamma^i_{jl}-
 \hbox{\Large$\frac{1}{2}$}\frac{\pd^2\rho}{\pd x^j\pd x^l})
 \theta^{jb}\frac{\pd}{\pd\theta_{la}}.
\end{eqnarray}
The function $\rho$ satisfies the relations
\beq
\label{M}
\pd_i\omega^{ij} = \hbox{\Large$\frac{1}{2}$}\omega^{ji}\frac{\pd\rho}{\pd x^i}.
\eeq
In particular this means that
\beq
\hbox{\Large$\frac{1}{2}$}\frac{\pd\rho}{\pd x^i}=\Gamma^j_{ij}.
\eeq
It is not difficult to solve the equation (\ref{M}) for $\rho$. Indeed,
it holds,
\begin{eqnarray}
 \frac{\pd\rho}{\pd x^k}=2\omega_{kj}\pd_i\omega^{ij}=
 (\omega_{kj}\pd_i\omega^{ij}+\omega_{ik}\pd_j\omega^{ij})=
 -\omega_{ji}\pd_k\omega^{ij},
\end{eqnarray}
and therefore
\begin{eqnarray}
\label{M1}
 \rho=-{\rm log}\;{\rm det}\;\omega^{ij}=
 {\rm log}\;{\rm det}\;\omega_{ij}
\end{eqnarray}
where the Jacobi identities in the form
$\omega_{kj}\pd_i\omega^{ij}+
\omega_{ik}\pd_j\omega^{ij}+\omega_{ji}\pd_k\omega^{ij}=0$ were used.

From (\ref{R}) the final expression for the anticommutator 
(\ref{DaVb}) follows:
\begin{eqnarray}
\label{DaVb1}
 \Delta^aV^b+V^b\Delta^a=\hbox{\Large$\frac{1}{2}$}\left( 
 \vep^{ab}\omega^{ij}R^k_{\;lji}\theta_{kc}\frac{\pd}{\pd\theta_{lc}} +
 R^k_{\;lij}\theta_{kc}\theta^{ib}\frac{\pd}{\pd\theta_{ja}}
 \frac{\pd}{\pd\theta_{lc}}+ R^k_{\;jkl}
 \theta^{jb}\frac{\pd}{\pd\theta_{la}}\right ) .
\end{eqnarray}

In its turn the actions of $\Delta^a$ and $V^a$ on the antibrackets 
$(\;,\;)^a$ are given by the rule 
\begin{eqnarray}
\nonumber
\Delta^{\{a}(f,g)^{b\}}- (\Delta^{\{a}f, g)^{b\}}
+(-1)^{\epsilon(f)}(f,\Delta^{\{a}g)^{b\}}=\\
=(-1)^{\epsilon(f)}\left({\hat R}^{ab}(fg)-({\hat R}^{ab}f)g-
f({\hat R}^{ab}g)\right),
\end{eqnarray}
\begin{eqnarray}
\nonumber
V^a(f,g)^b- (V^af, g)^b+(-1)^{\epsilon(f)}(f,V^ag)^b=\\
=\hbox{\Large$\frac{1}{2}$} R^k_{\;lij}\theta^{ia}\theta_{kc}\left(
\frac{\pd f}{\pd\theta_{lc}}\frac{\pd g}{\pd\theta_{jb}}-
\frac{\pd f}{\pd\theta_{jb}}
\frac{\pd g}{\pd\theta_{lc}}\right).     
\label{}
\end{eqnarray}

The properties of compatibility for $(\;,\;)^a$ have the form
\begin{eqnarray}
\nonumber
&&(-1)^{(\epsilon(f)+1)(\epsilon(h)+1)}(f,(g,h)^{\{a})^{b\}} + ~
{\mbox{\rm cycl. perm.}}~(f,g,h) = \\
\nonumber
&&=\left({\hat R}^{ab}(fgh)-
f{\hat R}^{ab}(gh)-{\hat R}^{ab}(fg)h-
{\hat R}^{ab}(fh)g(-1)^{\epsilon(h)\epsilon(g)}+\right.\\
&&\left.+({\hat R}^{ab}f)gh +f({\hat R}^{ab}g)h+
fg({\hat R}^{ab}h)
\right)(-1)^{\epsilon(f)\epsilon(h)+\epsilon(f)+\epsilon(g)+\epsilon(h)}.
\end{eqnarray}                                                                                
Therefore, on a flat Fedosov manifold, $R^k_{\;lij}=0$, an explicit 
realization of the strong triplectic algebra is constructed.

\pagebreak

\end{document}